\documentclass[preprint,aps,floatfix]{revtex4}
\usepackage{graphicx}
\usepackage{graphicx}
\usepackage{epsfig}
\usepackage{amssymb}

\newcommand{\beq}{\begin{equation}}
\newcommand{\eeq}{\end{equation}}
\newcommand{\bea}{\begin{eqnarray}}
\newcommand{\eea}{\end{eqnarray}}
\begin{document}
\preprint{\vbox{\hbox{JLAB-THY-07-612} }}
\vspace{0.5cm}
\title{\phantom{x}
\vspace{0.5cm} Strong and Electromagnetic Mass Splittings in Heavy Mesons}

\author{
J.~L.~Goity$^{~a b}$ 
\thanks{e-mail: goity@jlab.org} and
C.~P.~Jayalath$^{~a}$ 
\thanks{e-mail: jayalath@jlab.org}}

\affiliation{
$^{a}$Department of Physics, Hampton University, Hampton, VA 23668, USA. \\
and  $^b$Thomas Jefferson National Accelerator Facility, Newport News, VA 23606, USA. \\
}
\begin{abstract}
The contributions to heavy meson mass differences by the strong hyperfine interaction, the light quark masses and the electromagnetic interaction are
obtained from the empirical values of the $D$, $D^*$, $B$ and $B^*$ masses by means of a  mass formula based on the heavy quark mass expansion.
The three different types of contributions are determined with significant accuracy to next to leading order in  that expansion. 
\end{abstract}


\maketitle

\section{Introduction}

Hadron masses reveal key aspects of QCD. The light pseudoscalar meson masses reveal the spontaneous breaking of chiral symmetry as well as its explicit breaking by the light quark masses, and in a rather direct manner they permit to extract the ratios of these quark masses which are fundamental parameters of QCD. They also give access to the electromagnetic contributions to the masses, for instance,    in the mass difference
between charged and neutral pions. In the framework of chiral perturbation theory,  a rather accurate understanding of the effects of light quark masses and electromagnetic corrections on the pseudoscalar octet has been achieved, and to a lesser extent in baryons as well.  Heavy mesons represent another kind of hadronic system where one can arrive at a good determination of the various effects that determine their masses. Also, to a first degree of approximation in the heavy quark expansion,  the strong hyperfine effects,  which  involve the heavy quark spin,  serve to determine the ratio $m_c/m_b$,  which is another fundamental input in QCD.  In this work,  we show that the current knowledge of the heavy meson masses allows for  a quantitative determination of the different effects that contribute  to the differences of heavy meson masses. The approach followed here is similar to the one given by Rosner and Wise \cite{RosnerWise},  and the improvement in the results is possible thanks to the better empirical accuracy in the heavy meson masses and also in the better knowledge of the heavy quark masses.   

In the limit of infinite heavy quark masses and of light quark $SU(3)$ symmetry, heavy mesons fill multiplets of $U(2N_F) \times U(3) \times SU_j (2)$, where $N_F=2$ is the number of heavy quarks, $SU_j(2)$ is the rotation group associated with the light degrees of freedom of the heavy meson.  In particular, the ground state mesons, namely $D$,  $D^*$,  $B$ and $B^*$,  fill the multiplet  $(4,\;3,\;2)$ of that symmetry.  The symmetry is broken by several effects:
\begin{itemize}
\item The finite masses of the heavy \emph{c} and \emph{b} quarks break $U(4) \times SU_j(2) \to SU_J(2) \times U_c(1) \times U_b(1)$, where   $SU_J(2)$  is the rotation group  is associated  with the spin of the meson.
\item The light quark masses break $U(3) \to U_u(1) \times U_d(1) \times U_s(1)$.
\item The EM interactions break all symmetry subgroups down to $ SU_J(2) \times U_c(1) \times U_b(1) \times U_u(1) \times U_d(1) \times U_s(1)$.
\end{itemize}
The analysis in this work has the objective of sorting out  these three sources of symmetry breaking from the current knowledge of the heavy meson masses.
  The  analysis only involves meson mass differences. Those in the $D$ mesons are very well known with   errors smaller than 2.5\% \cite{PDG}. In the  $B$ mesons,  not all mass splittings are established; the mass differences   $(B^{*-}-B^-)$ and    $(B^{*0}-B^0)$  are not separately known  and,  in addition,  the errors are significantly larger than in the D-system  \cite{PDG}. Nonetheless, the available information is sufficient for the analysis to lead to  significant conclusions.
  
The analysis is based on the   mass formuli  that result from the expansions  in $1/m_Q$, where $m_Q$ is the heavy quark   mass $(Q = c, b)$, in $m_u, m_d$ and $ m_s$,  and in the fine structure constant $\alpha$.  These mass formuli   are  similar to the ones given long ago by Rosner and Wise \cite{RosnerWise}, except that the QCD running of some of the parameters are taken into account.  Using the notation $H = M_H$ , etc., they read as follows:
\begin{equation}
 H^{(*)}  = H_0^{(*)} (m_Q ) + \kappa ^{(*)} (m_Q )\; m_q  
 + \alpha\; (a^{(*)} (m_Q )\;{\mathbf Q}_q^2  - b^{(*)} (m_Q )\;{\mathbf Q}_Q \;{\mathbf Q}_q ),
\end{equation}
where $q$ and $Q$ denote respectively  the light and heavy quark flavors of the heavy meson, the label  $*$ is used for  the vector mesons, and ${\mathbf Q}_q$ and ${\mathbf Q}_Q$ are the respectively the light and heavy quark charges.

The first term in the mass formula is made out of the contributions in the limit $m_q\to 0$ and $\alpha\to 0$ plus $SU(3)$ singlet contributions by light quark masses, {\it i.e.} given in terms of $m_u+m_d+m_s$, and by electromagnetism. These $SU(3)$ symmetric contributions are simply absorbed into the coefficients that determine $H_0^{(*)} (m_Q )$ in Eqn.(2) below. Our analysis   cannot determine those contributions separately.
 The second term in the mass formula provides the $SU(3)$ breaking effects by the quark masses to first order, and the last two terms provide the $SU(3)$ breaking by the electromagnetic interaction, the first one representing the electromagnetic self-energy of the light antiquark,  and the second one is the electromagnetic interaction between light antiquark and  heavy quark. Throughout, $m_Q$ represents the heavy quark  pole mass. 
Up to ${\cal{O}}(1/m_Q^2)$, the $1/m_Q$  expansion gives for $H_0^*(m_Q)$:
\begin{equation}
 H_0^* (m_Q ) = H_0 (m_Q ) + h_1 (m_Q )\;\frac{{m_\rho ^2 }}{{m_Q }} 
  + h_2(m_Q) \; \frac{{m_\rho ^3 }}{{m_Q^2 }} . 
\end{equation}
For dimensional purposes, we use the $\rho$-meson mass as the reference  mass scale.
The explicitly displayed terms ${\cal{O}}(1/m_Q )$ and higher represent the mass difference produced by the strong hyperfine interaction. $h_1(m_Q)$ has a non-trivial dependence on $m_Q$ that results from the QCD running 
of the effective heavy quark  operator in the $1/m_Q$ expansion associated with it.
The heavy quark effective Lagrangian in the $1/m_Q$ expansion  is given in standard notation by \cite{EichtenHill}:
  \begin{equation}
  {\cal L}_{eff}  = \overline Q _v\, iv\cdot D\,Q_v  - \frac{{1}}{{2m_Q }}\;\overline Q _v\; D^2 Q_v  \\
+ g_s\,\frac{{C_{M}(\mu)  }}{{4m_Q }}\;\overline Q _v \,\sigma _{\mu \nu } \,G^{_{\mu \nu } } \,Q_v  + O(1/m_Q^2 ),
\end{equation}
where  the third term,  which couples to the heavy quark spin,  gives rise to $h_1$.  The coefficient $C_M(\mu)$,  where $\mu$ is the renormalization scale, is given by \cite{EichtenHill}:
\begin{equation}
C_M (\mu ) = C_M(\mu _0 )\left(\frac{{\alpha _s (\mu )}}{{\alpha _s (\mu _0 )}}\right)^{-\frac{C_A}{\beta_0}}
\end{equation}
with the initial condition given by the  matching to full QCD \cite{EichtenHill}:
\begin{equation}
C_M (m_Q ) = 1 + (C_A  + C_F )\;\frac{{\alpha _s (m_Q )}}{{2\pi }}.
\end{equation}
Here  $C_A = 3$ , $C_F = 4/3$ , and $\beta_0 = 11 - \frac 23 N_f$ is the first coefficient of the $\beta$-function with $N_f$ the number of flavors lighter than $m_Q$. 
Because  $h_1$ is proportional to $C_M$, we can express it in the following renormalization group invariant form:
\begin{equation}
h_1 (m_Q ) = C_M (m_Q )\; \alpha _s (m_Q )^{\frac{C_A}{\beta _0 }}\,\; \bar h _1, 
\end{equation}
where  $\bar h _1 $ is $m_Q$ and $\mu$ independent. In what follows,  all over-lined coefficients are  $m_Q$ and $\mu$ independent. 
The term  proportional to $h_2$ receives  contributions from two terms in the heavy quark Lagrangian at  ${\cal{O}}(1/m_Q^2)$ \cite{Balzereit}.  Taking into account the running of $h_2$ with $m_Q$ is thus impossible in this analysis. Fortunately,  this is not important because   the ${\cal{O}}(1/m_Q^2 )$ terms play a minor role  in the $B$ mesons, and thus neglecting the running of the $h_2$  is a good approximation.  

For the light quark mass effects,   $m_q(\mu)$ is defined in $\overline {MS}$ scheme and one has:
\begin{eqnarray}
 \kappa (m_Q ,\mu ) &=& \kappa _0 (\mu) + \kappa _1 (\mu )\;\frac{{m_\rho  }}{{m_Q }}
- \frac{3}{4}\;\kappa '_1 (m_Q ,\mu )\;\frac{{m_\rho  }}{{m_Q }} ~~,\nonumber\\
 \kappa ^* (m_Q ,\mu ) &=& \kappa _0 (\mu ) + \kappa _1 (\mu )\;\frac{{m_\rho  }}{{m_Q }} 
 + \frac{1}{4}\;\kappa '_1 (m_Q ,\mu )\;\frac{{m_\rho  }}{{m_Q }} ~.
 \end{eqnarray}
Because the spin-independent term in the ${\cal{O}}(1/m_Q)$ heavy quark Lagrangian is scale independent,  $\kappa_0$ and $\kappa_1$ are independent of $m_Q$ and their dependence on $\mu$ is given by the  running of $m_q(\mu)$, {\it i.e.}
 $ \kappa _i (\mu ) = \alpha _s (\mu )^{ - \frac{4}{\beta_0}}\; \bar \kappa  _i $, $i=0,1$. On the other hand,   because $\kappa'_1$   is  proportional to $C_M$,  it has an extra running factor similar to that of $h_1$:
 \begin{equation}
\kappa '_1 (m_Q ,\mu ) = 
C_M (m_Q )\;
\alpha _s (m_Q )^{\frac{C_A } {\beta _0 }} \;\alpha _s (\mu )^{-\frac{ 4}{\beta _0 }}\; \bar {\kappa}_1{'} .
\end{equation}
The $\mu$ dependence given above is rather immaterial in our analysis,  where the same $\mu$ is used for $D$ and $B$ mesons.

In the case of electromegnetic  effects, we need to discuss seperately the two terms. 
The coefficients of the self-energy   can be expressed in the following most general form:  \begin{eqnarray}
a(m_Q ) &= &  \bar{a}_0  +  \bar{a}_1\; \frac{{m_\rho  }}{{m_Q }}-\frac{3}{4}\,a_1'(m_Q) \; \frac{{m_\rho  }}{{m_Q }}\nonumber\\
 a^* (m_Q )& =&  \bar{a}_0  +  \bar{a}_1\; \frac{{m_\rho  }}{{m_Q }}+\frac{1}{4}\,a_1'(m_Q) \; \frac{{m_\rho  }}{{m_Q }}.
\end{eqnarray}
We note that  $a_1'$  will run with $m_Q$ in a similar form as $h_1$.  The spin of the heavy quark will affect very little the  self-energy term, and therefore   $a_1'$ will be small.   As shown later, $a_1'$ can be eliminated because of linear dependencies, which means that the effect associated with it cannot be distinguished from other effects on the meson masses. 

  On the other hand, the electromagnetic  terms  involving  the interaction between the light degrees of freedom and the heavy quark have the form:
\begin{eqnarray}
b(m_Q ) = \bar{b}_0  + \bar{b}_1\; \frac{{m_\rho  }}{{m_Q }} - \frac{3}{4}\;b'_1 (m_Q )\;\frac{{m_\rho  }}{{m_Q }}\nonumber\\
b^* (m_Q ) = \bar{b}_0  + \bar{b}_1\; \frac{{m_\rho  }}{{m_Q }} + \frac{1}{4}\;b'_1 (m_Q )\;\frac{{m_\rho  }}{{m_Q }}
\end{eqnarray}
$\bar{b}_0$ and $\bar{b}_1$  give the leading and subleading in $1/m_Q$ contributions to the Coulomb  interaction,  and both are  scale independent  as it  is known from  the  renormalization of  current operators of the heavy quark, such as the electromagnetic current.   The electromagnetic  hyperfine effect proportional to   $b'_1$  can be expressed as follows: 
\begin{equation}
b'_1 (m_Q ,\alpha _s (\mu )) = \bar{b}_1^{(1)}  + C_M (m_Q )\;\alpha _s (m_Q )^{\frac{C_A }{\beta _0 }} \;\,\bar{b}_1^{(2)}.
\end{equation}
It receives  contributions from  two general types of  diagrams  shown in  Fig (1). The  first term  results from  the coupling of the photon to the heavy quark spin and therefore it has no $m_Q$ dependence, while the second term
corresponds to the coupling of a gluon to the heavy quark spin,  and is therefore proportional to $C_M$ and thus $m_Q$ dependent.

\section{Analysis}
In this  analysis, we consider the five different mass splittings possible in each multiplet, {\it i.e.} $(D^+-D^0)$, $(D^{*+  }  - D^+)$, $(D_s-D^0)$, $(D^{* 0  }  -D^0)$, $(D_s^*-D_s)$ and similarly for B-system.
The mass formuli leave one parameter independent mass relation, which reads:
\begin{equation}
  ((B_s^*  - B_s ) - (B^{*0 }  - B^0 )) = \frac{ m_c \,\chi (m_b )} {m_b\, \chi (m_c )}\;((D_s^*  - D_s )-(D^{*+  }  - D^ +  ) ) ,
\end{equation}
where we denote 
$ \chi (m_Q ) \equiv C_M (m_Q )\;\alpha _s (m_Q )^{\frac{C_A }{\beta _0 }}$. This mass relation  is violated   by terms ${\cal{O}}(1/m_Q^3)$, ${\cal{O}}(\alpha /m_Q^2)$, ${\cal{O}}(m_q^{3/2})$, and  ${\cal{O}}(m_q /m_Q^2)$. Note that this relation was discovered in Ref.\cite{RosnerWise},  where the evolution factor $\chi(m_Q)$ was not included.

If one disregards the term ${\cal{O}}(1/m_Q^2)$ in the strong hyperfine interaction, {\it i.e.} the term proportional to $h_2$, one obtains an additional relation:
\begin{equation}
\frac{ m_c\; \chi (m_b )\;((D^{*0}  - D^0 ) + 2\,(D_s^*  - D_s ))}  
  {m_b\; \chi (m_c )\;((B^{* -  }  - B^ -  ) + 2\,(B_s^*  - B_s ))} = 1 .
\end{equation} 
The  deviations from  this relation  are  a measure of the importance of the $1/m_Q^2$ term  in the  hyperfine  interaction.

In the mass formuli,  there is  a total of twelve parameters  that enter in mass differences.   Since there are   ten mass differences and one parameter free mass relation,   there must be three linearly dependent terms in the mass formuli that we can eliminate.  The linear dependencies are such that $\bar{a}_1$ and $\bar{b}_1$ can be absorbed into  $\bar{a}_0$ and $\bar{b}_0$, and  $a_1'(m_Q)$  into $h_1(m_Q)$ and $b'_1(m_Q)$. Since no dependencies appear if one  stays at leading order in $1/m_Q$, it is natural to eliminate sub-leading terms. We, therefore,  eliminate therefore  $\bar{a}_1$, $a'_1$ and $\bar{b}_1$. The linear dependencies imply    that,   in this analysis,  one  cannot    determine  the $1/m_Q$ corrections to the  self-energy  and to the  Coulomb effects independently from   other effects.   In what follows,  we therefore set:
\begin{equation}
\bar{a}_1=a_1'(m_Q)=\bar{b}_1=0.
\end{equation}

The quark masses are the key input parameters in the mass formuli. If we would only keep up to ${\cal{O}}(1/m_Q)$ terms, the ratio $(D^*-D)/(B^*-B)$  would determine $m_c/m_b=0.40$ for $m_b\sim 5$ GeV.  There are more accurate determinations of the ratio $m_c/m_b$ from the analysis of charmonium and bottomonium \cite{Eidemuller} that give $m_c/m_b=0.35\pm0.03$ and $m_b=4.98\pm 0.13$ GeV. With this ratio,  it becomes necessary to include the ${\cal{O}}(1/m_Q^2)$ term in Eqn. (2). Note that Eqn. (12) could be used to extract the ratio $m_c/m_b$, but it requires a precision  in the mass differences involved that is far beyond the current precision. For the light quark mass inputs,  we  only need their ratios, namely, $m_s/\hat{m}$ where $\hat{m}=(m_u+m_d)/2$, and $m_s/(m_d-m_u)$, which  have been extensively studied in chiral perturbation theory \cite{Leutwyler}. The first ratio is obtained  from the ratio $M_K^2/M_\pi^2$,  which after next to leading order chiral corrections,  gives  a value of $24.4\pm 1.5$,  and the second ratio requires the input of isospin breaking observables, in particular the mass ratio
$(M_{K^0}-M_{K^+})/M_K^2$, and the rates for $\eta\to 3\pi$, with a result $m_s/(m_d-m_u)=42.5\pm 3.2$
\cite{Leutwyler}. For further reference, this latter ratio corresponds to having the electromagnetic mass difference $(K^+-K^0)_{\rm EM}=2.0\pm0.4$ MeV. The final input is $\Lambda_{QCD}$ required by $\alpha_s$; we use $\Lambda_{QCD}=200$ MeV.

With our inputs for the heavy quark masses,  we  obtain   for the  left hand side of  Eqn. (13) a value equal to $0.90\pm0.08$, which gives some evidence  for the need of the  ${\cal{O}}(1/m_Q^2)$ term in Eqn. (2).  For Eqn. (13) to hold,   it would be necessary to have $m_c/m_b=0.40$.  From Eqn. (12) we obtain  the combination:
\begin{equation}
(B^{*0}-B^0)-(B^{*}_s-B_s)=-0.90\pm 0.16~{\rm MeV},
\end{equation}
which is not  known experimentally  because the mass difference  $(B^{*0}-B^0)$ has not been  established  separately from the one for  charged ones. The improvement over similar prediction given in \cite{RosnerWise} is primarily due to the improved accuracy of the various inputs, especially the heavy quark masses and   the running effect  characterized by the factor $\chi(m_Q)$.

In Table I,  we give the results of our fits, displaying the partial contributions, and in Table II,  the effects are combined.  For the strong hyperfine contributions,  we see that the   ${\cal{O}}(1/m_Q^2)$  effect tends to reduce the  contribution from the  leading   ${\cal{O}}(1/m_Q)$  term by  up to 20\% in  D-mesons and by  up to 8\% in  B-mesons.  The errors quoted for the individual terms are rather large which indicate a strong correlation.  This is displayed in Fig.(2).  The error due to the uncertainty in $m_c/m_b$ manifests itself in the individual  strong hyperfine terms where it is approximately  $\pm 30$MeV in the D mesons. This indicates that,  at present, the  ${\cal{O}}(1/m_Q^2)$ strong hyperfine
effects cannot be established with good precision. Clearly,  the combined hyperfine effects are very precise as shown in Table II.

The leading order contributions by  the light quark masses,  which  are independent of the meson's spin, {\it i.e.}  the contributions proportional to $\kappa_0$, are well determined by the fit,    the error being dominated by the errors in the input light quark mass ratios.  The  ${\cal{O}}(1/m_Q)$  corrections  proportional to $\kappa_1$ can  also be  determined quite easily:  because the spin dependent term   $\kappa_2$  turns out to be rather small, the $\kappa_1$ term is almost entirely determined in terms of the    combination $(D_s-D^+)-(B_s-B^0)$. In particular,  its sign is positive because  $(D_s-D^+)>(B_s-B^0)$.  This  is a puzzling fact:   if one tries to interpret the $\kappa_1$ term from the point of view of a non-relativistic constituent quark model,  the   dependence of $(H_s-H)$ on the light quark masses  will be through the reduced mass of the $\bar{q}{Q}$ system,  and because the ${\cal{O}}(m_q/m_Q)$ correction to the reduced mass is negative,  the sign of the $\kappa_1$  term should be opposite to that of the $\kappa_0$ term. This is indication  of a clear   departure from the non-relativistic  quark model picture which we have not seen addressed in the literature.     To complete the light quark mass effects,   the terms proportional to $\kappa_2$  provide  the $SU(3)$ breaking effects  to the strong hyperfine interaction , and are determined with an error of about 10\%, which is remarkable.  For instance,  in the case of the $(D_s^*-D_s)$  mass difference,  it gives an upward shift of $3.3\pm 0.3$  MeV.  The combined effects of the light quark masses, displayed in Table II,  show that they are determined   by this analysis with
an accuracy that is in general  better than 10\%.
There is,  however,
one important and still unresolved problem concerning the light quark mass effects, and this has to do with the non-analytic contributions  proportional to  ${\cal{O}}(m_q^{3/2})$ \cite{GoityChPTHM}, which  are expected to be large \cite{GoityChPTHM,Randall},  according to the estimated value  of the coupling $g$ \cite{GoityRoberts} that gives the amplitudes  $D^*\to D\pi$.   A consistent analysis should include up to the $m_q^2$ effects, which is beyond the current analysis.  This problem, therefore,  introduces some uncertainty in the determination of the  light quark mass effects that is difficult to estimate. 

As mentioned earlier, we can separate the electromagnetic effects into self-energy  and Coulomb plus hyperfine type terms.  The self-energy  has only a spin   independent piece  and it represents an effect of less than 1 MeV,  and is determined with about  30\% error.   It has the  same sign and comparable  magnitude 
to   results  from  calculations  based on the Cottingham formula for electromagnetic  mass shifts  in a VMD approximation \cite{GoityHou,GoityEM}.     The effect of the Coulomb interaction is given by the term proportional to $\bar{b}_0$ and its subleading piece proportional to $\bar{b}_1$, as explained earlier,  has been  absorbed into  other terms. The fit determines the Coulomb effect with an error of 11\%. 
 The electromagnetic hyperfine effects are associated with the two parameters   $\bar{b}_1^{(1)}$ and   $\bar{b}_1^{(2)}$.   The  input for $(B^*-B)$  does not differentiate between the neutral and charged mesons;  if it is identified with the neutral mesons,  then one of these  two parameters  can be eliminated as
 a consequence of the relation Eqn. (12). For this reason,   in the fits,  we carried out identifying $(B^*-B)=(B^{*-}-B^-+B^{*0}-B^0)/2$;  one finds a large correlation between these two parameters,  which requires 
that   we keep only one of them. We keep  $\bar{b}_1^{(1)}$, which amounts to ignoring the $m_Q$ dependence in  $\bar{b}_1^{(2)}$,  absorbing the rest of it into  $\bar{b}_1^{(1)}$.  Our  analysis  is, therefore, insensitive to the QCD running of the electromagnetic  hyperfine effects, which is not surprising.   We have moreover checked that our  results are almost insensitive to the interpretation of the input  $(B^*-B)$ as an arbitrary  combination of the charged and neutral mass differences. The  hyperfine effects  are significant in the D-mesons, for instance,   in the $(D^+-D^0)$ case,  it is about  60\% of the Coulomb effect. On the other hand,  in the B-system,  the hyperfine effects are much smaller than the experimental uncertainties in the mass differences. Comparison,   with the calculations in Ref.\cite{GoityEM},  shows  agreement with the results  in the   elastic approximation to the Cottingham formula using VMD. Note that the inelastic contributions in the Cottingham formula that correspond to the interaction  Coulomb and hyperfine terms are suppressed by $1/m_Q$ and,  therefore,  to the order we are working here,  they can be neglected.
  
  It is instructive to make some comparisons. The electromagnetic shifts for the pseudoscalar mesons
  are $(D^+-D^0)_{\rm EM}=2.3\pm0.2$ MeV, and $(B^--B^0)_{\rm EM}=1.9\pm0.2$ MeV, which are similar within errors to $(K^--K^0)_{\rm EM}=2.0\pm0.4$ MeV. This is however a bit of a coincidence, as it can be seen from the results obtained for the different contributions,  which are in the  Coulomb and hyperfine   cases very different in the $D$ and $B$ mesons. 
If we consider the vector mesons, we obtain  $(D^{*+}-D^{*0})_{\rm EM}=0.8 \pm 0.2$ MeV, and $(B^{*-}-B^{*0})_{\rm EM}=1.6\pm 0.2$ MeV, which we can compare with $(\rho^+-\rho^0)=0.7\pm 0.8$ MeV (note that,  as in the case of pions, the mass difference between charged and neutral $\rho$ meson is purely electromagnetic).  One important observation is that the uncertainties in the ratios of light quark masses only have a noticeable effect in the self-energy terms, while the uncertainty in $m_c/m_b$  leaves the results for the electromagnetic effects virtually unchanged.
The isospin  mass splittings due to $m_u$ and $m_d$ are:  $(D^+-D^0)_{m_q}=2.42\pm 0.12$ MeV,
 $(D^{*+}-D^{*0})_{m_q}=2.50\pm 0.12$ MeV, $(B^0-B^-)_{m_q}=2.16\pm 0.2 $ MeV,  and   $(B^{*0}-B^{*-})_{m_q}=2.18\pm 0.2$ MeV.  The difference between the last two is the negligible isospin breaking induced on the strong hyperfine interaction in the $B$ mesons. This effect is larger and significant in the $D$ mesons as shown by the difference between the first two mass differences.  Finally, one can give an accurate   prediction: 
  $(B^{*0}-B^{*-})=0.6\pm0.2$ MeV.
 The  combination,  $\frac{3}{4}(\frac{1}{3}(D^+-D^0)+(D^{*+}-D^{*0})-\frac{1}{3}(B^0-B^-)-(B^{*0}-B^{*-}))_{m_q}=0.3\pm 0.02$ MeV,  gives the difference between $D$ and $B$ mesons of the  ${\cal{O}}(1/m_Q)$  spin independent part of the isospin breaking by the quark masses.  The electromagnetic part of the same combination is substantially larger and equal to $2.8\pm 0.2$ MeV.  A similar analysis is straightforward for the mass splittings due to $m_s$.

In summary, we have analyzed the different contributions to the mass splittings in  heavy ground state mesons. The analysis shows that,  with the current  empirical accuracy of the heavy meson masses,  one can determine these contributions with significant precision  even at the sub-leading order in $1/m_Q$.   
The results obtained here can be useful for  constraining  models  of heavy mesons, and perhaps also  for   lattice QCD calculations of heavy mesons  masses  where  it is possible to study the light quark mass dependence.
\vspace*{-5mm}

\acknowledgements
\vspace*{-5mm}
We would like to thank  N. Brambilla, J. Soto and A. Vairo for  useful discussions on heavy quark masses, and J. Soto for comments on the manuscript.   This work was supported by the Department of Energy  through contract DE-AC05-84ER40150 (JLG), and by the National Science Foundation  through  grant  PHY-0300185 and PHY-0555559.


\newpage

\squeezetable
\begin{tiny}
\begin{table}[h]
\begin{center}
\begin{tabular}{|c|c|c|c|c|c|c|c|c|}
\hline\hline
  	 $\Delta M$ 		 &  $\bar{h}_1$ 	& $\bar{h}_2$  	&  $\kappa_0$  		&   $\kappa_1$  	&  $\kappa_1'$  	&  $a_0$  		&    $\bar{b}_0$  		&  $\bar{b}_1^{(1)}$   \\
\hline

$D^+-D^0$  	&  0 			&0  			&$2.01\pm0.14$ 	&$0.47\pm 0.06$	&$-0.06\pm 0.01$&$-0.7\pm 0.2$	&$1.88\pm0.20$  	&$1.17\pm0.06$ 	 \\ \hline

 $D_s-D^+$ 	&   0 	 		& 0 			&$82.2\pm 2.8$	&$19\pm 2.3$ 		&$-2.3\pm0.2$  	& 0	 		& 0  				&  0 				  \\ \hline

$D^{*0}-D^0$ 	&$175\pm 29$	&$-34\pm 28$	&  0 				&  0 				&$0.09\pm 0.01$ 	&  0			&   0				&$1.0\pm0.05$		   \\ \hline

$D^{*+}-D^+$ 	&$175\pm 29$	&$-34\pm 28$	&  0 				&  0 				&$0.17\pm 0.01$ 	&  0			&   0				&$-0.52\pm0.03$	 \\ \hline

 $D_s^*-D_s$ 	&$175\pm 29$	&$-34\pm 28$	& 0  				&   0				&$3.3\pm0.28$ 		&  0			&   0				&$-0.52\pm0.03$	 \\ \hline

$B^0-B^-$ 	&  0			&  0   		&$2.01\pm0.14$	&$0.16\pm0.03$ 	&$-0.016\pm0.002$  &$-0.7\pm 0.25$&$-0.94\pm0.1$ 	&$-0.20\pm0.02$	   \\ \hline

 $B^*-B$ 		&$49.8\pm 3.5$&$-4.1\pm 3.3$&  0 				&  0 				&$0.038\pm0.005$ 	& 0 			&   0				&$-0.05\pm0.01$	\\ \hline

$B_s-B$ 		&0			&  0   		&$83.2\pm 2.9$	&$6.8\pm1.2$		&$-0.68\pm0.08$	&$-0.36\pm0.1$&$-0.47\pm0.05$	&$-0.10\pm0.01$	  \\ \hline

$B_s^*-B_s$ 	&$49.8\pm 3.3$&$-4.1\pm 3.3$& 0    			&  0 				&$0.94\pm0.13$ 	&  0 			&    0				&$0.09\pm0.01$	
\\
\hline \hline
\end{tabular}
\end{center}
\caption{  The mass shifts due to the different terms  are labeled by the corresponding coefficient in the mass formuli and are in the in units of MeV. The errors include the uncertainties in the input quark mass ratios. In the fit $(B^*-B)$ is interpreted as $\frac{1}{2}(B^{*-}-B^-+B^{*0}-B^0)$.}
\end{table}
\end{tiny}
\vspace{2cm}

\squeezetable
\begin{tiny}
\begin{table}[h]
\begin{center}
\begin{tabular}{|c|c|c|c|c|c|}
\hline\hline
  	 $\Delta M$ 		 &  Strong HF 		&    		Light quark masses	& Electromagnetic		& Total 			& PDG \cite{PDG}  \\
\hline
	
$D^+-D^0$  	&    $ 0   $ 			&  $  2.42\pm0.12   $ 	&  $ 2.33 \pm0.22   $ 	&$4.75\pm0.15$  	&   $4.78\pm 0.10$ \\ \hline

 $D_s-D^+$ 	&     $ 0  $ 			&   $  98.96\pm0.49   $ 	 &  $ 0  $ 		&$98.96\pm0.47$	 &  $98.85\pm 0.30$  \\ \hline

$D^{*0}-D^0$ 	&  $140.99  \pm0.1   $ 	&  $ 0.09 \pm0.01   $ 	&  $  1.04\pm0.05   $ 	&$142.12\pm0.1$ 	& $142.12\pm 0.07$   \\ \hline

$D^{*+}-D^+$ 	&  $ 140.99  \pm0.1   $ 	&  $  0.17 \pm0.01   $ 	&  $  -0.52\pm0.03   $  	&$140.64\pm0.09$ 	&  $140.64\pm 0.10$ \\ \hline

 $D_s^*-D_s$ 	&  $ 140.99  \pm0.1   $ 	&  $  3.30\pm0.28   $ 	&  $ -0.52 \pm0.03   $ 		&$143.77\pm0.3$ 	&   $143.8\pm 0.4$ \\ \hline

$B^0-B^-$ 	&    $ 0   $ 			&  $ 2.16 \pm0.12   $ 	&  $ -1.86 \pm0.19   $ 		&$0.304\pm0.11$  	&  $0.33\pm 0.28$   \\ \hline

 $B^*-B$ 		&  $45.70  \pm0.02   $ 	&  $0.04  \pm0.004   $ 	&  $  -0.05\pm0.002   $ 		&$45.69\pm0.02$  	&    $45.78\pm 0.35$\\ \hline

$B_s-B$ 		&  $ 0   $ 			&  $ 89.32 \pm0.37   $ 	&  $ -0.93 \pm0.09   $ 			&$88.39\pm0.33$ 	&$88.3\pm 1.8$  \\ \hline

$B_s^*-B_s$ 	&  $45.70  \pm0.02   $ 	&  $ 0.94 \pm0.08   $ 	&  $ 0.09 \pm0.005   $ 		&$46.73\pm0.08$  	& $45.3\pm 1.5$
\\
\hline \hline
\end{tabular}
\end{center}
\caption{  Mass contributions by strong hyperfine, light quark masses and electromagnetism in  units of MeV. The errors include the uncertainties in the quark mass ratios.   The fit has $\chi^2\sim 1$.
}

\end{table}
\end{tiny}
\vspace{5cm}
\newpage

\begin{figure}[h]
\includegraphics[width=110mm]{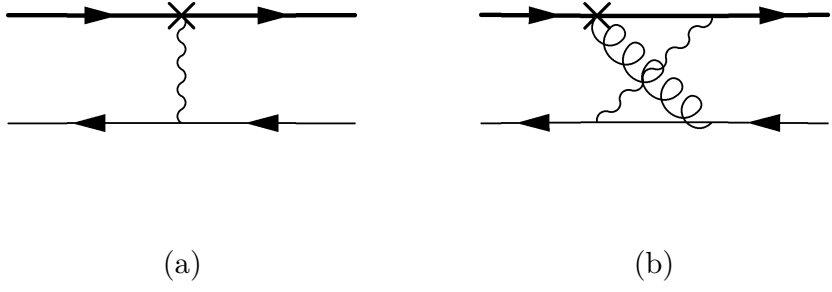}
\caption{Diagrams contributing to the electromagnetic mass shift parameters: type (a) contributes to $\bar{b}_1^{(1)}$,  and  type (b) to $\bar{b}_1^{(2)}$.
The crosses  indicate  respectively the coupling of the photon and the gluon to the heavy quark spin.}
\end{figure}

\vspace*{2cm}

\begin{figure}[h]
\includegraphics[width=130mm,height=90mm]{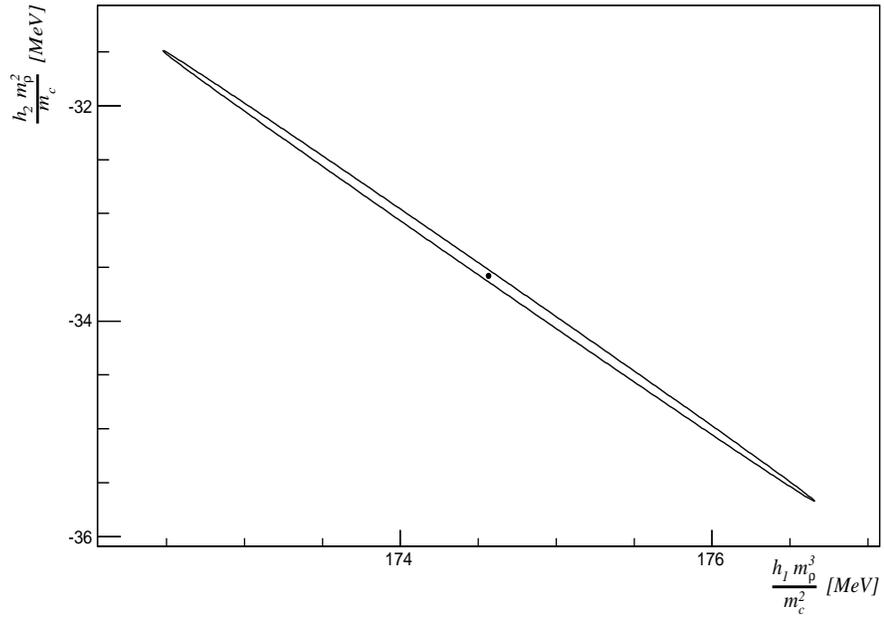}
\caption{Statistical correlation between the leading and sub-leading hyperfine contributions in $D$ mesons for  the fit with $m_c/m_b=0.35$.}
\end{figure}

\end{document}